\documentclass[twocolumn, twocolappendix]{aastex701}
\hypersetup{linkcolor=black,citecolor=blue,filecolor=cyan,urlcolor=blue, colorlinks=false}
\usepackage{xcolor}
\usepackage{mhchem}
\usepackage{booktabs}
\usepackage{xspace}

\newcommand{\kcalmol}{\ensuremath{\mathrm{kcal\,mol^{-1}}}}
\newcommand{\cxny}{\ce{C_{X}H_{Y}}\xspace}
\usepackage{newtxtext,newtxmath}
\newcommand{\rev}[1]{#1}
\received{February 1, 2025}
\revised{March 1, 2025}
\accepted{\today}
\graphicspath{{./}{figures/}}
\begin{document}

\title{Can cyanide radicals drive molecular backbone growth on interstellar icy grains?}

\author[orcid=0000-0001-8803-8684,sname='Molpeceres']{Germán Molpeceres}
\affiliation{Departamento de Astrofísica Molecular, Instituto de Física Fundamental, CSIC, C/ Serrano 123,113bis,121, E-28006 Madrid, Spain}
\email[show]{german.molpeceres@iff.csic.es}  

\author[orcid=0000-0002-2147-7735,sname='Enrique-Romero']{Joan Enrique-Romero}
\affiliation{Leiden Institute of Chemistry, Gorlaeus Laboratories, Leiden University, PO Box 9502, 2300 RA Leiden, The Netherlands}
\email[show]{j.enrique.romero@lic.leidenuniv.nl}

\begin{abstract}

Motivated by the value of CN-bearing molecules as tracers of interstellar physical conditions, we investigate the reactions of adsorbed CN radicals with acetylene and ethylene (\ce{C2H2} and \ce{C2H4}) on interstellar dust-grain analogues using quantum chemical calculations. We find that reactivity is strongly controlled by the relative orientation of the reactants. We further show that, on ice, these reactions differ qualitatively from their gas-phase counterparts, stalling at the formation of the adduct complexes \ce{C2H2CN} and \ce{C2H4CN} and exhibiting newly emerged kinetic barriers for the neutral-radical association. We contextualize our calculations in the same reaction-diffusion framework that would be employed in astrochemical models, finding that, depending on the diffusion energy of the hydrocarbons, these reactions can be either negligible or efficient, highlighting the importance of the local ice structure in interstellar grain chemistry. These findings caution against the use of CN-based tracers that assume barrierless, bimolecular surface reactions involving CN radicals.

\end{abstract}

%\keywords{\uat{Astrochemistry (75)}}


\section{Introduction} 

Cold interstellar gas-phase chemistry relies extensively on the barrierless addition of small radicals to unsaturated hydrocarbons, a reaction class that drives the rapid emergence of molecular complexity. In this context, the sequential growth enabled by CN or \ce{C2H} resembles a loose form of carbon-chain extension, which we refer to here as pseudohomologation. Unlike true homologation in terrestrial organic chemistry, this process does not follow a strict progression within a defined homologous series, but instead reflects a distinctive pathway of molecular growth that is unique to interstellar environments. The two simplest interstellar unsaturated hydrocarbons, which may contribute to their prevalence in the ISM, are acetylene (\ce{C2H2}) and ethylene (\ce{C2H4}) \cite{Barr2020ApJ,% high mass
Nickerson2023ApJ,%high mass
vGelder2024,%low mass
Arabhavi2024Sci}%disks
\xspace \rev{which are also thought to be abundant on interstellar icy grains, but have not yet been clearly detected, owing to the overlap of their characteristic bands with strong features of water and silicates, (as shown, e.g., by \citealt{1998A&A...331..749B} for \ce{C2H2})}. Despite of this, \ce{C2H2} and \ce{C2H4} have been suggested to be readily formed on interstellar ices by energetic processing of \ce{CH4}-rich ices \citep[e.g.,][]{Bennett2006ApJ,mifsud2023proton,Carrascosa2020MNRAS}. Notice that \ce{C2H2} and \ce{C2H4} have been detected in the comae of comets \citet[e.g.,][]{MummaCharnley2011,DRusso2016,DRusso2022}, with abundances of about 0.1-0.2 \% with respect to water, and also \textit{in situ} on the 67P comet \cite{Hanni2022}. \rev{However, experimental and theoretical studies also provide evidence that these species react on interstellar ice surfaces \citep{2000ApJ...532.1029H,Kobayashi2017,molpeceres_radical_2022}. Such processes contribute to their destruction, but at the same time suggest that these hydrocarbons must be relatively abundant if their reaction products are commonly observed \citep{molpeceres_radical_2022}. To our knowledge, astrochemical models rarely address the abundance of simple hydrocarbons explicitly, although they are often implicitly assumed to be abundant \citep{garrod_three-phase_2013}.} Finally, it must be noted though, that these molecules are not directly desorbed in a pristine way from the nuclei of comets, but its a mixture of this effect plus the destruction of other species, likely polymers.

In the gas phase, both \ce{C2H2} and \ce{C2H4} react with CN via barrierless addition \cite{balucani_formation_2000, balucani_crossed_2000, huang_crossed_1999,kaiser_formation_2012}, where the main reaction channels are

\begin{align}
\ce{C2H2 + CN &-> HC3N + H}, \label{reac:1} \\
\ce{C2H4 + CN &-> CH2CHCN + H}, \label{reac:2}
\end{align}

leading to the formation of the important astromolecules cyanoacetylene (\ce{HC3N}) and vinyl cyanide (\ce{CH2CHCN}). These same reactions are commonly included in grain-surface chemical models by analogy with the gas phase, that is, they are assumed to be barrierless and to produce a bimolecular product. Indeed, \citet{duan_alma_2025} identify reaction \ref{reac:1} on grains as the dominant route for the formation of \ce{HC3N} under hot-core conditions. However, the behavior of the CN radical on dust grains has recently been shown to be far from understood \citep{enrique-romero_complex_2024}, and any significant deviation from the gas-phase behaviour of reactions \ref{reac:1} and \ref{reac:2} carries important implications. For example, \ce{HC3N} is often used as a chemical clock in star-forming regions \citep{wang_simulations_2025}, and variations in its abundance can lead to misinterpretations of the physical and chemical evolution of these environments.

\begin{figure}
\centering
\includegraphics[width=\linewidth]{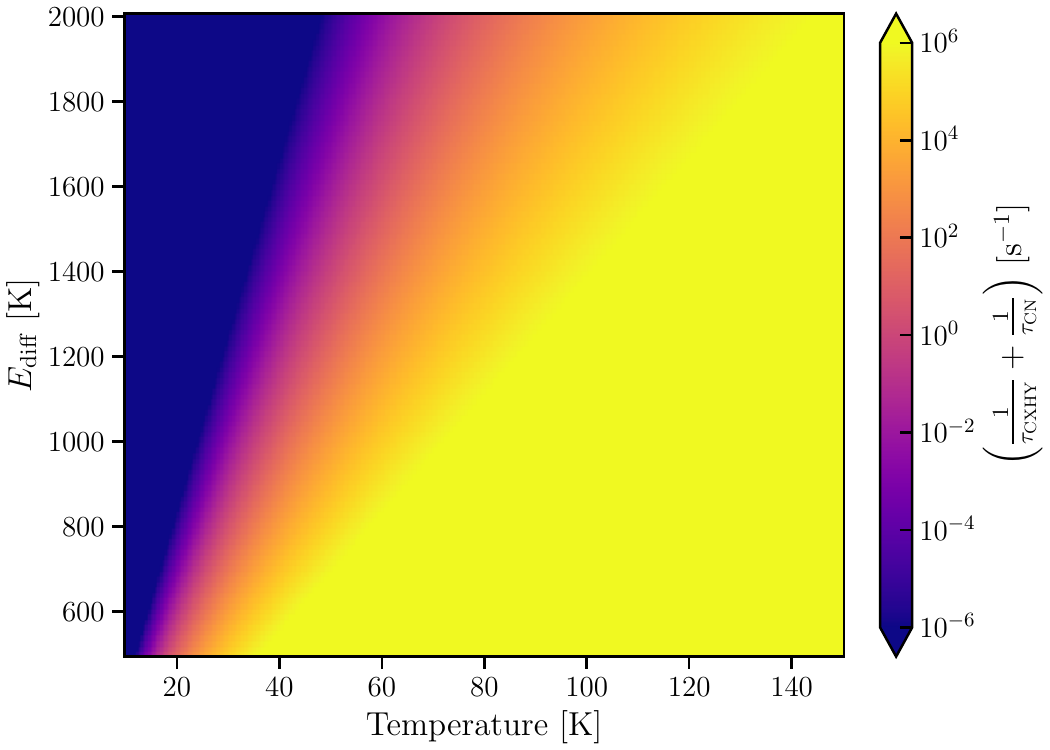}
\caption{Inverse of diffusion characteristic times for \ce{C_{X}H_{Y}} and CN, shown as a function of dust temperature and diffusion energy.  The color scale is saturated outside the range $10^{-6}$–$10^{6}$ s$^{-1}$; values represented with the extreme colors may therefore correspond to characteristic times significantly smaller or larger than these limits. An attempt frequency of $10^{12}$ s$^{-1}$ is assumed \citep{hasegawa_three-phase_1993}.}
\label{fig:Diffusion_rate}
\end{figure}

As mentioned above, the behaviour of the CN radical on ices is peculiar. In particular, \citet{enrique-romero_complex_2024} showed that CN radicals can form hemibonded complexes with water molecules, a type of non-classical bond in which an unpaired electron is shared between two atoms, in this case oxygen (from the surface water) and carbon (from CN). The hemibonded state of CN on ices is the most stable and abundant configuration \citep{enrique-romero_complex_2024}, and therefore governs the physicochemical behaviour of adsorbed CN. The expression for the Langmuir-Hinshelwood diffusive chemistry of reactions \ref{reac:1} and \ref{reac:2} \citep{hasegawa_three-phase_1993} may be summarised as

\begin{equation} \label{eq:lh}
k_{\rm LH} \propto \epsilon_{r} \left( \dfrac{1}{\tau_{\rm C_{X}H_{Y}}} + \dfrac{1}{\tau_{\rm CN}} \right) \sim \dfrac{\epsilon_{r}}{\tau_{\rm C_{X}H_{Y}}},
\end{equation}
where $\tau_{\rm C_{X}H_{Y}}$ is the characteristic diffusion time of the hydrocarbon and $\epsilon_{r}$ (also referred to as $\kappa$) is the reaction efficiency, defined as

\begin{equation}
\label{eq:efficiency}
\resizebox{\linewidth}{!}{$
\epsilon_{r} =
\dfrac{k_{r}}{k_{r} + k_{\rm diff, C_{X}H_{Y}} + k_{\rm diff, CN}
+ k_{\rm des, C_{X}H_{Y}} + k_{\rm des, CN} + k_{\rm \ce{H2O}}}
$},
\end{equation}
in reactions that present an activation barrier \citep{chang_gas-grain_2007}. The terms in the denominator account for the rates of reaction, diffusion, and desorption of each species, $k_{CN}$ is the specific contribution of CN; and $k_{\rm \ce{H2O}}$ describes the reactivity of CN with water to form HNCOH \citep{Rimola2018}. The binding energies that control the diffusion rates of \ce{C2H2} and \ce{C2H4} are not known, but estimates based on the \citet{Wakelam2017} approximation suggest values of approximately 2500 K for both molecules. A fraction of these energies corresponds to the diffusion barriers. In Figure \ref{fig:Diffusion_rate} we plot the characteristic diffusion times for reactions \ref{reac:1} and \ref{reac:2}, showing that any appreciable diffusive reactivity would only become efficient once the dust temperature begins to rise, at least above 40 K, from Fig \ref{fig:Diffusion_rate}. It is therefore clear that these reactions, if they take place at all, would occur during the warm-up phase of star-forming regions.

Thermal diffusion of the hydrocarbon is the first of the two intrinsic conditions required for any grain-surface reaction to occur. The second condition isa high reaction efficiency ($\epsilon_{r}$ in Equation \ref{eq:efficiency}). If the reaction were truly barrierless, $\epsilon_{r}$ would equal to 1, but if any activation energy exists, the reaction must compete with diffusion and desorption in general, and, for CN, also with chemisorption to form HNCOH. In this work we derive the temperature and diffusion energy dependence of $k_{\rm LH}$ for reactions \ref{reac:1} and \ref{reac:2}, determine the actual products formed on ices, highlight the unique role of hemibonded CN in the pseudohomologation of unsaturated hydrocarbons on interstellar grains, and discuss how our findings refine the astrochemical modelling of star-forming regions.

\section{Computational methods}

The electronic structure calculations supporting our results rely on previous protocols by us for the study of CN adsorption on \ce{H2O} ice. In brief, we start from a CN-14\ce{H2O} molecular cluster \citep{Molpeceres2021c,enrique-romero_complex_2024} that allow us to use a sufficiently accurate level of theory to describe the local CN-\ce{H2O} hemibonded interaction. This level of theory was already benchmarked in \citet{enrique-romero_complex_2024} and corresponds to DFT calculations using the M06-2X(D30) functional \citep{zhao_m06_2008, grimme_2010_d3} (\rev{where M06-2X refers to the core of the exchange and correlation functional and D30 refers to the correction to improve dispersion interaction}) and the def2-TZVPD basis set \citep{Weigend2005, rappoport_property-optimized_2010}. \rev{With this method, in \citet{enrique-romero_complex_2024} we showed an excellent agreement for the description of the hemibonded binding energy, as well as reaction barriers against high-level reference methods.} All DFT calculations were carried out with a large integration grid (\texttt{defgrid3} in \textsc{Orca}) to minimize the numerical noise, \rev{relevant to gradient and second derivatives calculations}, characteristic of the M06 suite of functionals. This method was used for geometry optimizations and molecular Hessian calculation. In addition, to encompass a proper treatment of the CN-\ce{C_{X}H_{Y}} interactions, particularly at the transition states (TS), all the electronic energies of the stationary points found in our search were refined using DLPNO-CCSD(T)/aug-cc-pVTZ \citep{purvis_full_1982,guo_communication_2018, Woon1994}. The protocol to find the transition states (TS) involved a relaxed PES scan defining a simple bond formation coordinate from preadsorbed states, followed by a Nudged Elastic Band (NEB) calculation \citep{henkelman200} taking the endpoints and the maxima along the reaction coordinate and optimization of the climbing image of the NEB calculation using standard geometry optimizers. Some reactions required all the steps and for some others the NEB step could be omitted as the maximum in the PES scan was very clearly defined. From the TS, we propagate the eigenvector of the reaction mode in both reaction directions to obtain the reactant and product state, from where we extract the activation energies ($\Delta U_{a}$) and reaction energies ($\Delta U_{r}$) reported in the main work. All the electronic structure calculations use the \textsc{Orca}(v6.0.0) code \citep{Neese2012,neese_orca_2020,neese_software_2025}.  Geometry optimizations were driven indistinctively with either the internal \textsc{Orca} optimizer or with an interface to the \textsc{Dl-find}/\textsc{Chemshell} engine \citep{Kastner2009,chemshell}

The reaction rate constants ($k_{r}$) used in the evaluation of equation \ref{eq:efficiency}, (also used for $k_{\ce{H2O}}$ in that equation) were obtained using Rice-Rampsberger-Kassel-Marcus (RRKM) theory for which we first calculate the microcanonical rate constant following Eq:
\begin{equation}
    k(E) = \Gamma \dfrac{N^{\ddagger}}{h \rho(E)}
    \label{eq:k_rrkm}
\end{equation}
where the sum and density of states of the transition state and reactants need to be calculated ($N^{\ddagger}$ and $\rho(E)$, respectively) and $\Gamma$ is the tunneling coefficient that is obtained assuming an assymetric Eckart barrier. Because our reactions take place atop ices, we assume infinitely fast energy dissipation on the water matrix, and therefore the microcanonical rate constants can be Boltzmann averaged according to:
\begin{equation} \label{eq:k}
    k_{\rm r} = k_{\infty}\left(T\right) = \dfrac{\int^{\infty}_{0} k(E)\rho(E)\exp^{-\beta E}{\rm d} E}{\int^{\infty}_{0}\rho(E)\exp^{-\beta E}{\rm d} E},
\end{equation}
where $\beta$=1/k$_{\rm B}$T and $\infty$ is used to represent the rate constant at \emph{high pressure}, that in our context should not be confused with actual pressure but rather fast quenching of the reaction energy with the ice. Finally, all $k_{\rm r}$ are collated and incorporated into Equation \ref{eq:efficiency} 

\section{Results} \label{sec:floats}

\subsection{Reaction coordinate sampling. Energetics of the CN + C$_{\rm X}H_{\rm Y}$ reactions} \label{sec:results_ener}

\begin{figure*}
    \centering
    \includegraphics[width=0.8\linewidth]{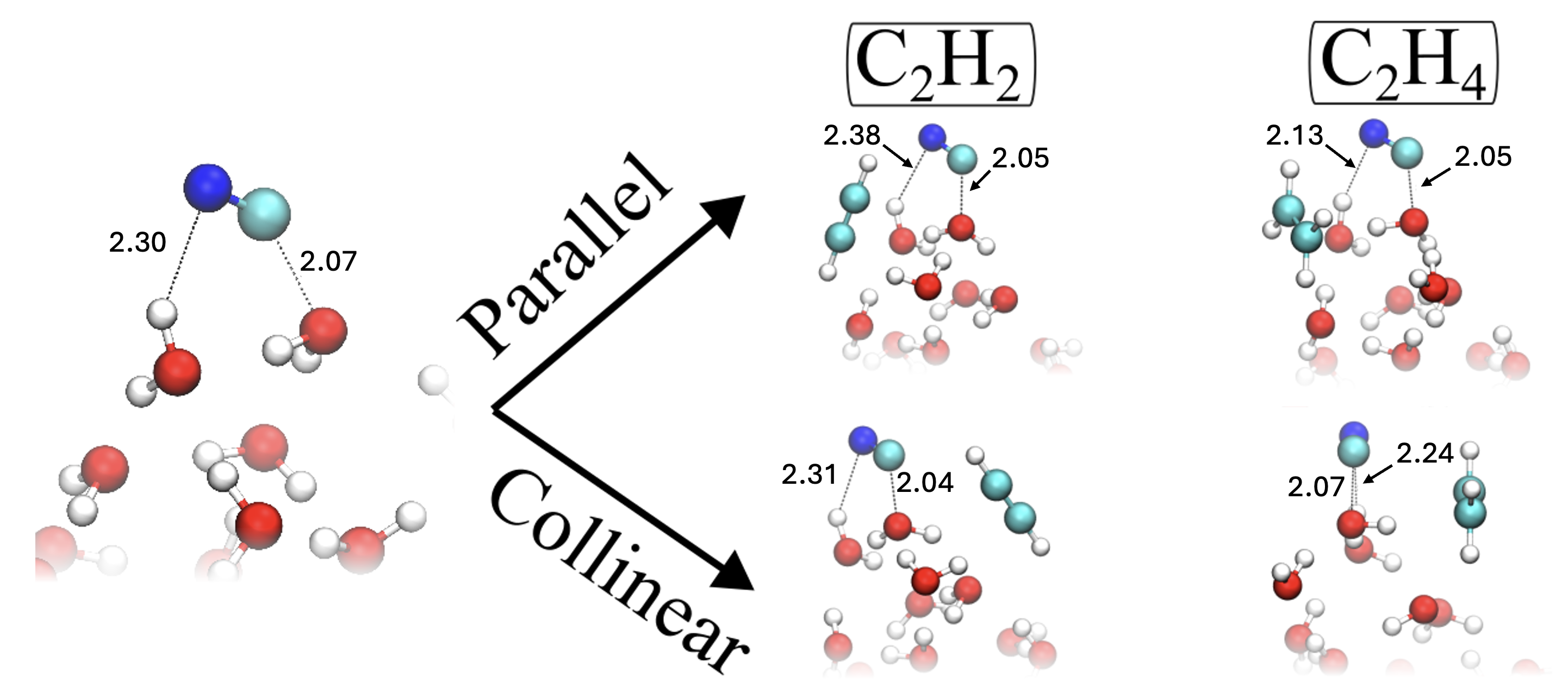}
    \caption{\rev{(Left) Depiction of the hemibonded CN radical adsorbed on a 14 \ce{H2O} cluster. (Right, Top) Parallel geometries for the approach of \ce{C2H2} and \ce{C2H4} to hemibonded CN. (Right, Bottom) Collinear geometries for the approach of \ce{C2H2} and \ce{C2H4} to hemibonded CN. All the structures are optimized and represent the actual reactant states used in the determination of the energetic descriptors of the reaction.}}
    \label{fig:structures_start}
\end{figure*}

\begin{deluxetable}{lccccc}
\tablecaption{Reaction descriptors for CN--hydrocarbon systems on water ice. $\Delta U^{\rm R}_{(1)}$ is the reaction energy for CN addition (the first step) forming the radical intermediate (\ce{C2H2CN} or \ce{C2H4CN}).
$\Delta U^{\ddagger}_{(1)}$ is the activation energy for the CN addition to the hydrocarbon.
$\Delta U^{\rm R}_{(2)}$ is the reaction energy for H elimination (the second step) yielding \ce{HC3N} or \ce{C2H3CN}.
$\Delta U^{\ddagger}_{(2)}$ is the activation energy for the H-abstraction step.
\label{tab:reaction_energetics}}
\tablehead{
\colhead{Reaction} &
\colhead{Orientation} &
\colhead{$\Delta U^{\rm R}_{(1)}$} &
\colhead{$\Delta U^{\ddagger}_{(1)}$} &
\colhead{$\Delta U^{\rm R}_{(2)}$} &
\colhead{$\Delta U^{\ddagger}_{(2)}$}
}
\startdata
\ce{CN + C2H2 -> C2H2CN -> HC3N + H} & Parallel   & -50.6 &  2.6  & 35.2 & 41.4 \\
\ce{CN + C2H2 -> C2H2CN -> H2CCCN}$^{a}$ & Collinear  & -52.6 &  2.2  & -6.3 & 38.9 \\
\ce{CN + C2H4 -> C2H4CN -> C2H3CN + H} & Parallel  & -52.7 & 3.1 & 33.2 & 37.5 \\
\ce{CN + C2H4 -> C2H4CN -> C2H3CN + H} & Collinear & -52.2 & 1.0 & 32.5 & 38.4 \\
\enddata
\tablecomments{
Energies are given in kcal~mol$^{-1}$ with respect to the CN/\cxny\ reactant state for each elementary step.
A positive reaction energy indicates that the corresponding step is endothermic, not that the overall process is endothermic.
Figure~\ref{fig:profiles} shows the reaction profiles using the reaction asymptote as the reference. $^{a}$ See text for the different products for the \ce{CN + C2H2} reaction.
}
\end{deluxetable}

\begin{figure*}
    \centering
    \includegraphics[width=\linewidth]{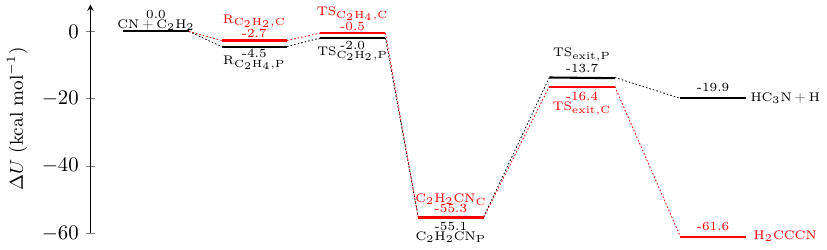} \\[2em]
    \includegraphics[width=\linewidth]{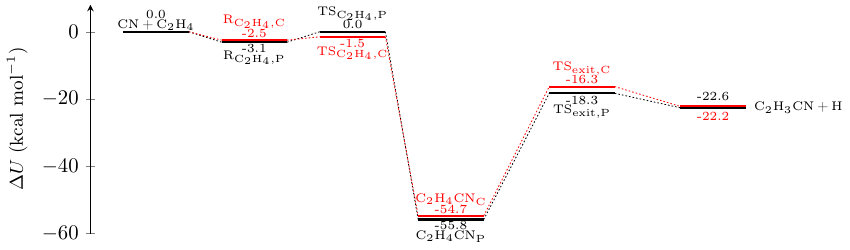}
    \caption{Reaction profiles for the CN + \cxny\ reaction in our ice analogues. (Top) \ce{CN + C2H2}. (Bottom) \ce{CN + C2H4}.}
    \label{fig:profiles}
\end{figure*}

The geometry of the hemibonded CN radical on the 14~\ce{H2O} cluster has already been reported in \citet{enrique-romero_complex_2024}. Nevertheless, it is worth showing it again here, as it not only constitutes the starting point of the present investigation but also illustrates the strong hemibond that characterizes the products of reactions~\ref{reac:1} and~\ref{reac:2}. The hemibonded radical is shown in Figure~\ref{fig:structures_start}. We also initiated our investigation from a hydrogen-bonded structure; however, as demonstrated in Appendix~\ref{app:HtoHemi}, its conversion into the hemibonded configuration involves a very small barrier, rendering the hydrogen-bonded motif kinetically irrelevant.

Starting from the geometry of the hemibonded CN radical, and assuming an anisotropic water-ice potential together with a random approach of the incoming molecules, we consider two distinct attack directions. The first, labelled ``parallel'', corresponds to the approach of the \cxny molecule with the C--C bond roughly equidistant to the CN axis. In the second approach, referred to in this work as ``collinear'', the \cxny molecule approaches with its C--C bond exclusively towards the carbon end of the CN radical. A schematic representation of these two attack directions is provided in Figure~\ref{fig:structures_start} where we show the geometry of the two pre-adsorbed molecules at the reactant state before \cxny\ diffusion. We often describe the motion in terms of the \cxny molecule rather than the CN radical, as the binding energy of the former is significantly smaller than that of the latter. Independent of the attack direction, both for \ce{C2H2} and \ce{C2H4} we find the \ce{C2H2CN} (\ce{HCCHCN}) and \ce{C2H4CN} (\ce{CH2CH2CN}) radicals as products. We also tested the possibility of a CN attack to the \cxny\ molecules where the CN comes from the gas-phase (a mechanism named Eley-Rideal). Such a mechanism is remarkably similar to the gas phase, and therefore, further discussion is given in Appendix \ref{app:er}.

By sampling these two attack directions, our goal is to assess whether the relative orientation of the reactants influences the breaking of the CN--\ce{H2O} hemibond, a necessary step for the reaction to proceed. This effect constitutes the key distinction with respect to gas-phase chemistry. An exploration of the reaction coordinate for CN addition starting from the adsorbed geometries shown in Figure~\ref{fig:profiles} unambiguously reveals, in all cases, the emergence of a kinetic barrier that is absent in the corresponding gas-phase reaction. All reactions are nevertheless found to be exothermic, as indicated by the negative values of $\Delta U^{\rm R}_{(1)}$.

The appearance of kinetic barriers on ices, in contrast to the barrierless behaviour in the gas phase, has been reported previously (see, for example, \citealt{enrique-romero_quantum_2022}). The activation energies $\Delta U^{\ddagger}_{(1)}$, reported in Table~\ref{tab:reaction_energetics} and Figure \ref{fig:profiles}, show some variability with respect to the attack direction. However, in all cases $\Delta U^{\ddagger}_{(1)}$ falls within the range of plausible diffusion barriers, $E_{\rm diff}$, for \cxny\ on water ice (Figure~\ref{fig:Diffusion_rate}), namely 1.0--3.1~\kcalmol\ (503--1569~K). Following Equation~\ref{eq:lh}, it is therefore legitimate to ask whether the emergent barrier can be effectively incorporated into the diffusion term, $\tau_{\rm C_{X}H_{Y}}$, that is, treated as a diffusion barrier, or whether its nature is instead purely electronic, in which case it should be included via Equation~\ref{eq:efficiency}. For reactions~\ref{reac:1} and~\ref{reac:2}, several lines of evidence support an electronic origin of the barrier. First, the presence of sizable kinetic barriers irrespective of the approach orientation indicates the absence of a random, diffusion-controlled component. Second, an analysis of the Löwdin partial charges at the reactant and transition states (Appendix \ref{app:c}) consistently shows an increased spin population localized on the carbon atom of the CN radical, rather than a delocalized distribution across the hemibonded atoms. This observation demonstrates that cleavage of the hemibond is a necessary step for reactions~\ref{reac:1} and~\ref{reac:2}, thereby ruling out diffusion as the origin of the barrier.
Taken together, these results support the conclusion that reactions~\ref{reac:1} and~\ref{reac:2} are not barrierless on water ice and must be modelled by explicitly accounting for an electronic activation barrier through Equation~\ref{eq:efficiency}.  Under those considerations we investigate the reaction efficiency later in Section \ref{sec:kinetic}. 

After CN addition to \cxny, a competition arises between thermalization of the newly formed radical (\ce{C2H2CN} or \ce{C2H4CN}) on the ice surface and the redistribution of excess energy into different molecular degrees of freedom. Among the possible non-thermal pathways, we have non-thermal diffusion, non-thermal desorption, and H-abstraction leading to the formation of \ce{HC3N} and \ce{C2H3CN}. Focusing on the latter process,  we have extended the investigation of Reactions~\ref{reac:1} and~\ref{reac:2}, which directly affects their chemical outcome. This corresponds to the rightmost parts of the reaction profiles shown in Figure~\ref{fig:profiles} and to the $\Delta U^{\rm R}_{(2)}$ and $\Delta U^{\ddagger}_{(2)}$ columns reported in Table~\ref{tab:reaction_energetics}.

The calculated H-abstraction activation energy barriers, $\Delta U^{\ddagger}_{(2)}$, consistently exceed 35.0~\kcalmol. This corresponds to more than 70\% of the the preceding reaction step (Table \ref{tab:reaction_energetics}), indicating that this process is highly unlikely compared with simple thermalisation of the addition complex. Such large activation barriers strongly suppress H-abstraction on the ice surface. Comparing with our previous studies on \ce{CO2} formation \citep{molpeceres_cracking_2023, ishibashi_proposed_2024}, we demonstrated both theoretically and experimentally that H-abstraction following the \ce{CO + OH} reaction is only marginally probable, leading predominantly to stabilization of the HOCO intermediate. This stems from its large abstraction barrier, in that case of approximately 30~\kcalmol. By analogy, we therefore conclude that the addition complexes \ce{C2H2CN} and \ce{C2H4CN} constitute the final products of Reactions~\ref{reac:1} and~\ref{reac:2} on water ice, rather than the bimolecular products \ce{HC3N + H} or \ce{C2H3CN + H}. The astrochemical implications of this result are discussed in Section~\ref{sec:astro}, and it represents yet another fundamental difference between gas-phase and grain-surface chemistry for these reactions.

Finally, although this step does not affect the overall reaction outcome, we find that H-abstraction from the \ce{C2H2CN} adduct in the collinear geometry does not yield the expected bimolecular products. Instead, it forms the \ce{H2CCCN} radical. 
This result can be rationalized by a favorable geometry for proton transfer mediated by surrounding water molecules \citep[see, e.g.][]{Rimola2018, Molpeceres2021c, Perrero2022, molpeceres_carbon_2024, bovolenta_-depth_2024}. Interestingly, \citet{baiano_gliding_2022} showed that the activation energy for proton transfer depends sensitively on both the number of molecules participating in the proton relay and the specific binding site. Our calculations represent a worst-case scenario, involving only two \ce{H2O} molecules in the proton-transfer motion. A more exhaustive exploration of alternative binding sites might therefore reveal conditions under which the formation of \ce{H2CCCN} is more favorable. Nevertheless, since the dominant outcome in all cases is thermalization of the addition complex, a detailed characterization of these secondary pathways lies beyond the scope of the present work.

\subsection{Reaction of hemibonded CN with \ce{H2O} ice} \label{sec:water}

\begin{figure}
    \centering
    \includegraphics[width=0.5\linewidth]{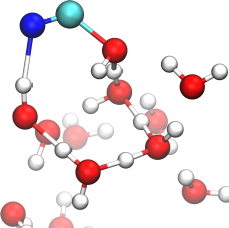}\\[2em]
    \includegraphics[width=0.9\linewidth]{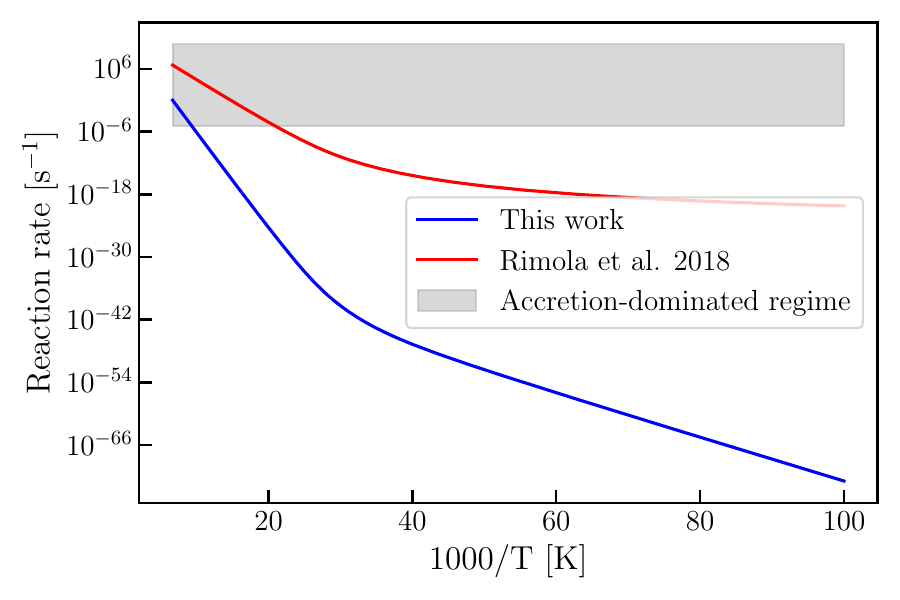} 
    \caption{(Top) Geometry of the transition state corresponding to the reaction with the water matrix in our setup (Bottom) Reaction rate constant for the reaction using our derived activation energy (8.5 \kcalmol) and \citet{Rimola2018} one (3.5 \kcalmol). We note that in the calculation of the rate constant only the activation energy was modified, with the vibrational and tunneling contributions coming exclusively from our calculations (see text). The gray box represent the H-accretion competing process, assumed to be 1 atom day$^{-1}$. The data used to generate this figure can be obtained as data behind the figure. }
    \label{fig:reaction_water}
\end{figure}

Before performing a kinetic analysis of the reactions considered in this work, we need to extract every possible term of equations \ref{eq:lh} and \ref{eq:efficiency}. In Section \ref{sec:kinetic} we enumerate how we calculate the diffusion and desorption terms. However, in the case of the reaction involving a hemibonded CN radical there is an additional competitive process, which is the reaction of the CN directly with the \ce{H2O} molecule to which it is hemibonded to. This is an activated process by $\sim$3.5 \kcalmol\ as reported by \citet{Rimola2018}, although it has a deeper influence of quantum tunneling owing to the motion of protons. The reaction with the water matrix can be formulated as
\begin{equation}
    \ce{CN + 2H2O -> HNCOH + H2O},
\end{equation}
while in reality in this type of reactions the whole \ce{H2O} matrix plays a role leading to a well-known variability in activation energies across binding sites and H-bond networks \citep{Molpeceres2021c,baiano_gliding_2022}. To gain context on the influence of the reaction with water in the whole reaction-competition scheme, we simulated the reaction at our specific level of theory while maximizing the number of partaking molecules in the reaction, as \citet{baiano_gliding_2022} report a correlation between the number of water molecules and the height of the barrier. From our search we locate a proton relay involving five water molecules as the maximum possible in our binding site. The activation energy from that binding site was found to be 8.5~\kcalmol\ which is 5~\kcalmol\ larger than the previously reported values. We cannot pinpoint the origin of this discrepancy, which may arise from variations in binding energies in the hemibonded geometry, steric constraints within the H-bond network, or differences in the theoretical method. In recent calculations on similar systems we managed to narrow down the range in activation energies for the reactions to 8.4--17.5~\kcalmol\ \citep{enrique-romero_lamberts_2026}, showing that the variability in values clearly indicates two limits of the possible activation energy distribution. 

The rate constants for the reaction with water, together with the optimized transition state geometry, are shown in Figure~\ref{fig:reaction_water}. Two sets of rate constants were computed. The first uses the activation energies derived in this work, while the second adopts the activation energy reported by \citet{Rimola2018}. In both cases, the vibrational and geometric properties of the transition state are those obtained in this work, so that the only difference between the two sets of rate constants is the activation energy. The values based on \citet{Rimola2018}, which yield systematically larger rate constants, are retained in Section~\ref{sec:kinetic}.

As can be seen in Figure~\ref{fig:reaction_water}, the rate constants are extremely small at low temperatures and only significant at relatively high temperatures, when thermal diffusion of every general adsorbate and reactivity in the ice are efficient. We therefore conclude that the reaction between CN and water can only proceed, especially at low temperatures, prior to full thermalization, for instance following a collision of CN with the ice surface. This interpretation is consistent with the microcanonical treatment of the rate constants presented by \citet{Rimola2018}.

Interestingly, the reaction with water ice constitutes, in some of our sampled cases, the most competitive term in Equation~\ref{eq:efficiency} at low temperatures when $E_{\rm diff}$ and $\Delta U^{\ddagger}_{(1)}$ are large, owing to the contribution of quantum tunneling in the CN + H$_2$O reaction and the lack of it for the other terms. Nevertheless, the overall Langmuir--Hinshelwood rate (Equation~\ref{eq:lh}) remains negligibly small because diffusion is inefficient. As a result, even a high reaction efficiency leads to a vanishing net production of reaction products, a point that we will revisit in the following section.

\subsection{Kinetic analysis: Reaction-diffusion competition} \label{sec:kinetic}

\begin{figure*}
    \centering
    \includegraphics[width=\linewidth]{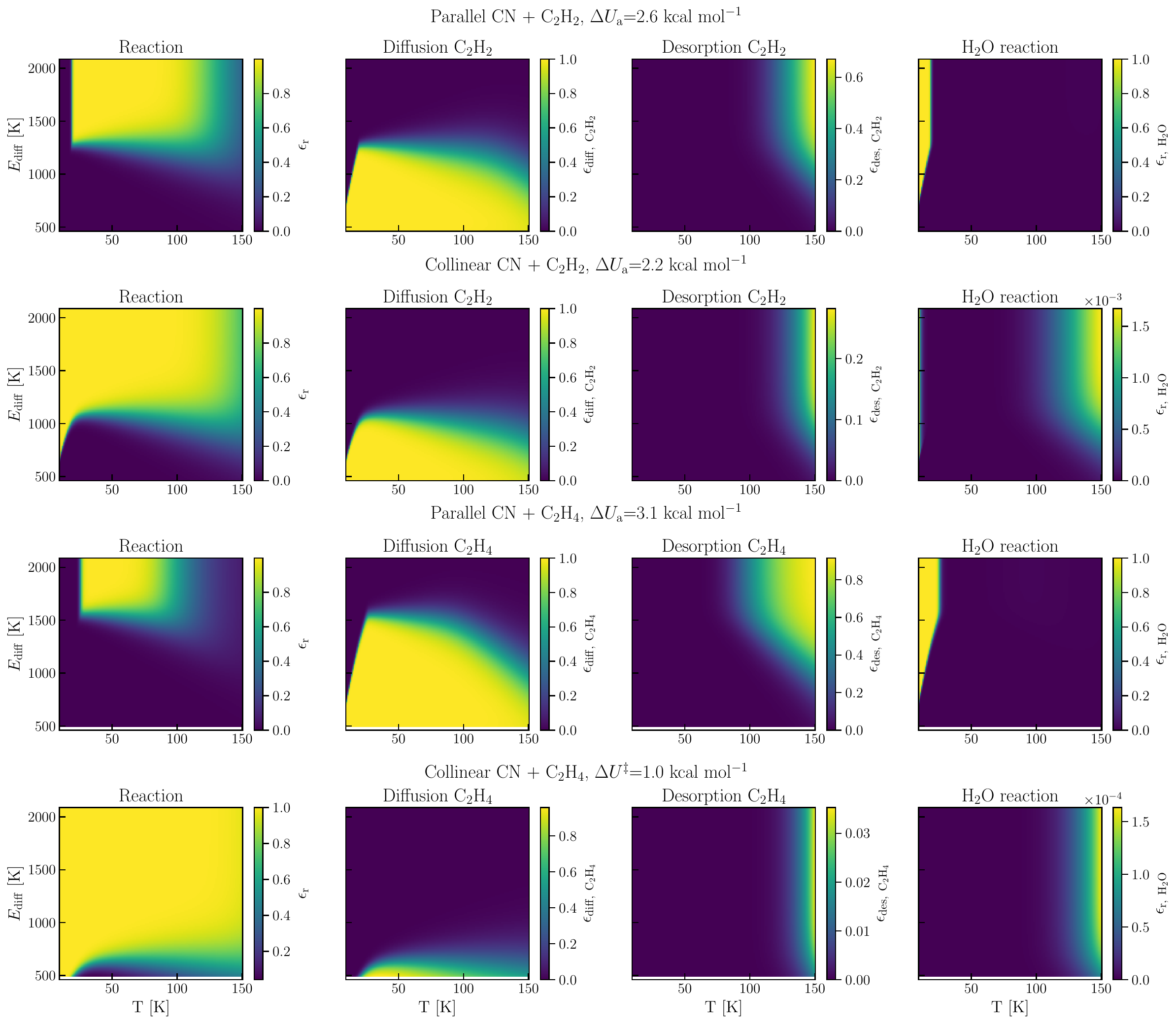} \\
    \caption{Reaction with \cxny\ (Top \ce{C2H2} and bottom \ce{C2H4}), diffusion, desorption and reaction with water efficiencies obtained as a function of $E_{\rm diff}$. Note that CN diffusion and desorption are neglected. The data to reproduce the leftmost column can be retrieved as data behind the figure.}
    \label{fig:competition}
\end{figure*}

Having examined the emergence of kinetic barriers in Reactions~\ref{reac:1} and~\ref{reac:2} in Section~\ref{sec:results_ener} (and in Appendix~\ref{app:c} to confirm their electronic origin), as well as the reaction of CN with \ce{H2O}, we can now assess the likelihood of forming \ce{C2H2CN} and \ce{C2H4CN} from Reactions~\ref{reac:1} and~\ref{reac:2}. The formation of \ce{HC3N} and \ce{C2H3CN} has already been excluded.

In order to apply Equation \ref{eq:efficiency}, we directly use the reaction rate constants for both the CN addition and the reaction with water, as obtained in Sections \ref{sec:results_ener} (and as data behind the figure of Figure \ref{fig:competition}) and \ref{sec:water}, respectively. \rev{For $k_{\ce{H2O}}$ we use the \citet{Rimola2018} activation energy as it is the lowest of the two considered ones, and it allows to gauge the impact of this factor in an optimistic scenario for the reaction with water.} For the desorption and diffusion terms, we adopt the following expressions:
\begin{align}
k_{\rm des,i} &= \nu_{\rm des,i}\exp\left(-\dfrac{\rm BE}{k_{\rm B}T}\right), \\
k_{\rm diff,i} &= \nu_{\rm diff,i}\exp\left(-\dfrac{E_{\rm diff}}{k_{\rm B}T}\right),
\end{align}
where $\nu_{\rm des,i}$ and $\nu_{\rm diff,i}$ are the pre-exponential factors for desorption and diffusion of CN and \cxny, BE is the binding energy, and $E_{\rm diff}$ is the diffusion energy, considered a factor 0.4 of BE. We also calculate diffusion, desorption and reaction with water competition efficiencies as part of our analysis, simply substituting the numerator in Equation \ref{eq:efficiency}. \rev{These competition terms are not strictly relevant in a modelling context, where diffusion, desorption, or \textit{on-site} non-diffusive reaction processes (reaction with water) are treated explicitly without a competition scheme like the one in Equation \ref{eq:lh}. They are nevertheless shown here because they illustrate which processes dominate during a CN+\cxny\ encounter and thus reduce the overall reaction efficiency.}

For $\nu_{\rm des,i}$, we adopt the value recommended by \citet{minissale_thermal_2022} for \ce{C2H2} (3$\times$10$^{16}$ s$^{-1}$). For \ce{C2H4}, which is not reported in \citet{minissale_thermal_2022}, we take the average of the values given for \ce{C2H2} and \ce{C2H6}, resulting in $\nu_{\rm des}=1.98\times10^{16}$ s$^{-1}$. For CN, we adopt the standard value of $1\times10^{12}$ s$^{-1}$, as CN desorption is negligible over the relevant kinetic timescales. We further assume a constant value of $\nu_{\rm diff}$ for all species involved. Binding energies are taken from \citet{Wakelam2017}, namely 2500 K for \ce{C2H4} and 2600 K for \ce{C2H2}. Although the binding energy of \cxny\ may differ and could be refined using more sophisticated approaches, we note that $E_{\rm diff}$, which is the most influential parameter in the application of Equation \ref{eq:efficiency}, is treated as a free parameter in our model. As a result, the precise value of the binding energy plays only a secondary role in the present analysis. Nevertheless, the explicit computation of the binding energy of \cxny\ is part of our ongoing work, given its relevance not only for the present reaction, but, more generally  for warm surface chemistry.

The results of our analysis are shown in Figure~\ref{fig:competition}, where they are interpreted independently of the \cxny\ species, as both molecules exhibit similar behavior. It is important to reiterate in this section that our analysis is framed in the contex reaction-diffusion competition \citep{chang_gas-grain_2007} and that, prior to reaction the two reactants must meet in a Langmuir-Hinshelwood picture, i.e., even if the reaction efficiency is 100\% the total rate constant portrayed in Equation \ref{eq:lh} might be negligible. A first key conclusion is that the relative orientation of the reactants plays a major role in determining reaction efficiency.

A difference of only 1.0 \kcalmol in the activation energies leads to either inhibit completely the reaction in favour of diffusion or to make it dominant over the entire temperature range. For example, in the \ce{CN + C2H4} reaction, the collinear orientation leads to reaction dominance across all temperatures, whereas in the parallel orientation, the reaction is largely suppressed, except within a narrow region at low temperatures and high E$_{\rm diff}$. \rev{All other outcomes, namely desorption and reaction with water, are negligible in the whole temperature range when competing with diffusion and reaction, except for desorption at high temperatures and reaction with water at very low ones.} The former is due to the larger pre-exponential factor for desorption relative to diffusion and latter to the role of quantum tunneling in the competition between the reaction of interest and reaction with water. \rev{Certainly,  desorption at elevated temperatures will occur at non-negligible (and even dominant) rates, leading to the desorption of \cxny. However, this corresponds to a different process within the model. Here we focus exclusively on the reaction–diffusion competition occurring during a single encounter. Reaction with water, on the other hand, despite dominating the reaction–diffusion competition, exhibits very low rate constants in all cases, of the order of 10$^{-21}$ s$^{-1}$ at 10 K (see Figure \ref{fig:reaction_water} for the Rimola values). Thus, although it becomes the fastest process in certain regions of the parameter space shown in Figure \ref{fig:competition}, it remains intrinsically extremely slow. This prompts us to consider non-thermal scenarios and inefficient energy dissipation as possible explanations for the apparent prevalence of this reaction, an issue that will be explored in future work. }

As mentioned previously, the precise diffusion against reaction behaviour strongly depends on $\Delta U^{\ddagger}_{(1)}$. However, in general, low diffusion energies (E$_{\rm diff}<$1000 K) combined with barriers of arroud 2~\kcalmol favour diffusion out of the reaction site, whereas higher diffusion energies favour reaction dominance. The clear competition between diffusion and reaction arises from the relatively low predicted binding energy (and hence E$_{\rm diff}$) of \cxny\ on water ice, a parameter that is not well constrained and that requires an in-depth investigation considering the limited contemporary estimations \citep{silva_interaction_1994,Wakelam2017, behmard_desorption_2019, bovolenta_-depth_2024,minissale_thermal_2022, bariosco_gaseous_2025}, especially in light of binding energy distributions rather than single values. Notably, for \ce{C2H2}, reported binding energies range from 2500--3500~K, depending on the methodological approach. 

Overall, our investigation of the kinetics of reactions \ref{reac:1} and \ref{reac:2}, including the emergence of kinetic barriers in the entrance and the inhibition of H-abstraction, point to a scenario in which both reactions is the formation of the \ce{C2H2CN} and \ce{C2H4CN} radicals on water ice, with reaction efficiencies significantly below 100\%. 

\section{Discussion} \label{sec:discussion}

\subsection{Integration of the results into astrochemical models} \label{sec:models}

A large portion of the quantum chemical simulations in astrochemistry is devoted to the improvement of 3-phase astrochemical models of interstellar environments. However, the incorporation of the calculated data into models is not always straightforward. This is especially true in the modelling of grain-surface chemistry, where a quantitative treatment of reaction requires the use of continuous rather than discrete parameters \citep{furuya_framework_2024}. To the well established distributions of binding energies \citep[just citing a few of different laboratories:][]{Molpeceres2020,bovolenta_binding_2022, sameera_ch_2021, 10.3389/fspas.2021.645243, tinacci_theoretical_2023, bovolenta_co_2025}. Although less numerous, there are also specific attempts to model binding site heterogeneity for surface diffusion \citep{kars_diffusion-desorption_2014, asg17, Zaverkin2021, bariosco_fully_2025}. In this work we extend this requirement to activation energies of electronic nature. The variability in activation energies was something already observed by \citet{sameera_systematic_2023}, but the nature of the transition state was not resolved in that study. 

The inclusion of multiple activation energies is particularly problematic because the functional form of Equation~\ref{eq:lh} hinges on a binary assumption: namely, whether the reaction is barrierless. Although the use of single values is sometimes convenient or unavoidable, the resulting rate constant must be described by a steep, piecewise-defined function of the form:
\[
k_{\rm LH} \propto
\begin{cases}
\alpha \left(\dfrac{1}{\tau_{\rm C_{X}H_{Y}}} + \dfrac{1}{\tau_{\rm CN}} \right)
& \text{if } \text{BL}, \\[6pt]
\alpha \epsilon_{r}
\left( \dfrac{1}{\tau_{\rm C_{X}H_{Y}}} + \dfrac{1}{\tau_{\rm CN}} \right)
& \text{if } \text{BM} .
\end{cases}
\]
where BL and BM denote \emph{barrierless} and \emph{barrier mediated}, respectively, and $\alpha$ is a constant for every reaction, normally the branching ratio. Expressions of this kind, although simple, are cumbersome to implement for the handful of exceptions in the astrochemical reaction networks. A much easier way to circumvent the existence of orientations with activation energy is to simply modify the branching ratio for the reaction $\alpha$ to account for the fraction of orientations that preclude reaction, based on analyses such as those presented in Section \ref{sec:kinetic}. Therefore, for the reactions studied in this work, our recommendation is simply to assume an $\alpha$ value between 0.75--0.25 depending on the expected $E_{\rm diff}$ of \cxny .

\subsection{Implications for CN driven astrochemistry} \label{sec:astro}

The molecules subject of this study \ce{C2H2} and \ce{C2H4} are the most abundant unsaturated hydrocarbons detected in solid bodies of solar, cometary or interstellar origin \cite{Barr2020ApJ,% high mass
Nickerson2023ApJ,%high mass
vGelder2024,%low mass
Arabhavi2024Sci, MummaCharnley2011,DRusso2016,DRusso2022, Hanni2022}. Owing to their role as molecular scaffolds for the formation of more complex organic species, the abundances of some related molecules, such as \ce{HC3N}, are commonly used to infer physical properties of the local medium where they are detected, like for example its evolutionary age. This work therefore serves as a cautionary note regarding the use of molecular abundances as probes of interstellar medium physics when the underlying chemistry is not fully constrained. In particular, we show that Reaction~\ref{reac:1} does not directly yield \ce{HC3N + H}, but instead forms \ce{C2H2CN}, whose most likely subsequent evolution (e.g., via hydrogenation, based on the location of the unpaired electron) is:

\begin{equation}
    \ce{C2H2CN + H -> CH2CHCN}
\end{equation}

although the H-abstraction to \ce{HC3N + H2} must be also possible. Nevertheless, it is evident that the chemistry leading to \ce{HC3N} is not fully understood. For example, only in 2025 its hydrogenation destruction routes were postulated \citep{raaphorst_toward_2025}. The same reasoning applies to Reaction \ref{reac:2}. 

Our results do not fully disqualify CN as a driver of grain-surface chemistry, because larger species with larger E$_{diff}$ will have a lower chance to diffuse away from the reaction center on a diffusion-reaction scheme. Because the activation energy of the reactions stem from breaking the hemibond it is expected that the reaction of \ce{CN} with larger hydrocarbons known to react in the gas phase without a barrier, e.g., \ce{C4H6}, \ce{C3H6}, \ce{C3H8} \citep{morales_experimental_2010} will have a comparable activation barrier for a larger E$_{diff}$. Nonetheless, the abundance of these latter hydrocarbons on grains is expected to be much lower which will limit the overall impact of CN-driven gas-grain chemistry, specially in comparison with the gas-phase where CN reactivity is one of the most important drivers of molecular complexity. 

%\textcolor{red}{Aqui Joan habla de su articulo de CN en hielo. Y de como solamente se puede formar formamida si el CN colisiona con exceso de energía que es lo que haremos}

The interaction of and reactivity of CN with interstellar ices are strongly influenced by the nature of the ice substrate and the associated energetic landscape. Quantum chemical studies by \citet{Rimola2018} showed that while direct reaction of HCN with water ice exhibits large activation barriers and is not viable under cold interstellar conditions, CN radicals can react with water molecules in the ice to initiate formamide (\ce{HCONH2}) via a sequence of water-assisted hydrogen-transfer steps, which we reproduced in Figure \ref{fig:reaction_water}. In this mechanism, water molecules are not just spectating, but they actively participate in the reaction mechanism, catalyzig the H-transfers lowering the barrier with respect to the intramolecular H-transfer case. As shown in \ref{sec:water}, this mechanism strongly depends on the reaction site and the local water structure, and in addition, requires some extra energy, e.g., in \citet{Rimola2018} it was assumed that the entirety of the hemibonded CN's binding energy on water could be employed to surmount the barrier with water leading to formamide. 

More recent work by \citet{enrique-romero_complex_2024}, expands on CN's surface chemistry on both \ce{H2O} and \ce{CO} ices. These authors find that CN physisorbs strongly on water ice, frequently forming hemibonded complexes, whereas on CO ice its interaction is weaker and often limited to van der Waals-type binding; this substrate dependence affects subsequent reactivity. On water ice, CN can readily react with H to to form HCN and HNC, meanwhile with \ce{H2}, only the HCN channel is accessible with a sizeable barrier which would require quantum tunnelling to actually work. These reactions destroy CN radicals on the surface. Therefore the presence of free roaming H atoms will limit the amount of CN able to react with water and other admolecules like other radicals \citep{ER2025}, or the ones studied here (\ce{C2H2} and \ce{C2H4}).  Overall, these studies further underscore how binding energy and local ice structure govern whether additional energy input or specific pathways are required for further reaction to occur.

%\begin{itemize}
%    \item Evidence for CN being a fundamental tracer of gas-phase organic chemistry
%    \item CN can still react in the ice, provided that is thermalized and that the second reactant is another radical (citamos a Joan segundo y primer articulo)
%    \item The path to formamide needs to be microcanonical
%\end{itemize}

\section{Conclusion}

In this work we have highlighted just another intricate effect of the complexity of grain-surface interstellar chemistry. In particular, we have found emerged barriers for the reaction of CN with hydrocarbons, absent in the gas phase. The barriers that we found cannot be classified as diffusion barriers, therefore affecting the functional form of the rate constants in astrochemical models. We also ruled out the possibility of forming \ce{HC3N} and \ce{C2H3CN} from reactions \ref{reac:1} and \ref{reac:2}, respectively, as H-abstraction from the addition complex is strongly inhibited on water ice. The final products of these reactions on water ice are therefore \ce{C2H2CN} and \ce{C2H4CN} radicals. The complete change in the reaction route with respect to the gas phase, together with contemporary advances in the study of cyanopolyine hydrogenation under cold conditions and the fate of CN on water ice \citep{raaphorst_toward_2025,enrique-romero_complex_2024}, serve as word of caution for the use of CN-driven grain-surface chemistry to interpret the abundance of nitrogen-bearing organics and hence on the use of CN derivatives as tracers of physical conditions in regions with a clear gas-phase / grain-surface chemistry interplay.
%% Please use the acknowledgment and contribution environments. This will 
%% be anonomyized when the "anonymous" style option is used. 
\begin{acknowledgments}
GM acknowledges support from ERC grant ISOCOSMOS, GA No. 101218790, funded by the European Union. G.M. also acknowledges the support of the grant RYC2022-035442-I funded by MCIU/AEI/10.13039/501100011033 and ESF+ and PID2024-156686NB-I00. G.M. also received support from project 20245AT016 (Proyectos Intramurales CSIC). We acknowledge the computational resources provided by the DRAGO computer cluster managedby SGAI-CSIC, and the Galician Supercomputing Center (CESGA). The supercomputer FinisTerrae III and its permanent data storage system have been funded by the Spanish Ministry of Science and Innovation, the Galician Government and the European Regional Development Fund (ERDF). 
J.E.R. acknowledges the support from the Horizon Europe Framework Programme (HORIZON) under the Marie Skłodowska-Curie grant agreement No 101149067, ``ICE-CN'', and the access to the HPC resources of the high-performance computer SNELLIUS, part of the SURF cooperative of educational and research institutions in The Netherlands, under project No EINF-6197. J.E.R also acknowledges funding from NWO XS: OCENW.XS25.3.125 grant.
\end{acknowledgments}

\software{\textsc{ChemShell} \citep{chemshell}, \textsc{ORCA(v6.0.0)} \citep{Neese2012,neese_orca_2020,neese_software_2025}}


\appendix

\section{Conversion of H-bond to hemibonded CN in a 14 \ce{H2O} water cluster} \label{app:HtoHemi}

\begin{figure}
    \centering
    \includegraphics[width=\linewidth]{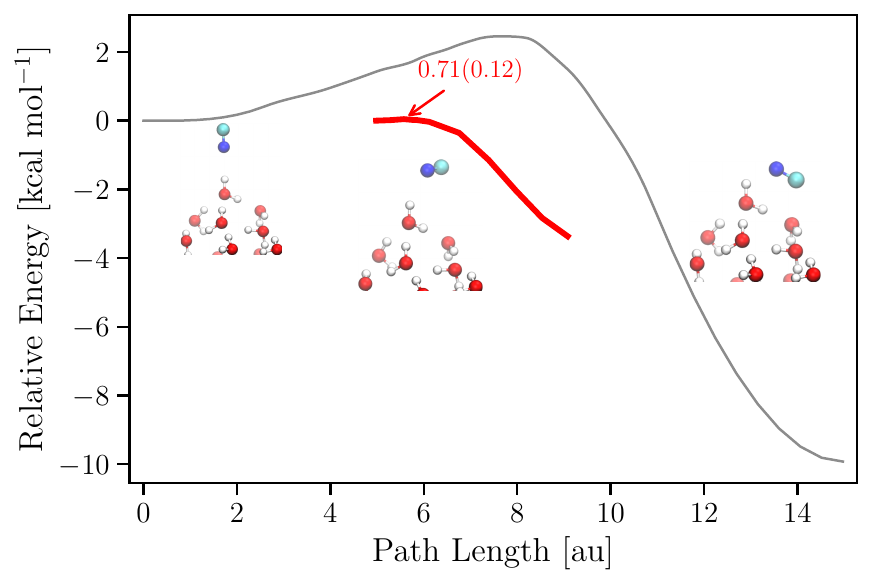}
    \caption{NEB profile for the H-bond to Hemibond states of CN radicals on the a binding site of the \ce{14H2O} cluster. In red, the final (ZOOM-NEB-CI) path after optimization of the zoomed region around the transition state. The text indicates the final barrier (in \kcalmol), including ZPE contributions at the DLPNO-CCSD(T)/aug-cc-pVTZ//M06-2X(D30)/def2-TZVPD level and at the M06-2X(D30)/def2-TZVPD one (in parenthesis).}
    \label{fig:neb_interconversion}
\end{figure}

The interpretation of the CN--\ce{H2O} binding proposed by \citet{enrique-romero_complex_2024} postulated two possible binding modes: a high-binding (more stable) energy hemibonded state, which is the configuration extensively considered throughout that work, and an additional lower-binding (less stable) energy state labeled the ``H-bond'' configuration, originating from the interaction of the N atom of CN with a dangling hydrogen bond of the water matrix. This latter mode was reported to have binding energies significantly lower than those of the hemibonded state.

The existence of the H-bond configuration was inferred from geometry optimisations, that is, without any explicit consideration of thermal effects, effectively corresponding to a 0~K picture. By contrast, the present work focuses on the effects associated with the warm-up phase during protostellar formation. A relevant question is therefore whether H-bond configurations become populated at intermediate or warm ice temperatures or whether, instead, the bimodal binding energy distribution reported by \citet{enrique-romero_complex_2024} collapses into a single distribution dominated exclusively by the hemibonded state.

We carried out NEB calculations to investigate the interconversion between the H-bonded and hemibonded adsorption states, starting from the H-bonded geometry reported in \citet{enrique-romero_complex_2024}. All calculations were performed at the same level of theory used throughout the main text, namely M06-2X-D3(0)/def2-TZVPD. The NEB profile obtained from 12 interpolated images is shown as a faded grey line in Figure~\ref{fig:neb_interconversion}. From this profile, an activation energy of approximately 2~\kcalmol\ can be inferred for the interconversion between the two states, indicating a relatively small barrier. To obtain a more accurate characterization of the transition region, the initial pathway was further refined using the ZOOM-NEB-CI protocol \citep[essentially a second NEB performed on a region zoomed around the maximum of the original path, with the highest-energy image released from string forces;][]{asgeirsson_nudged_2021}. This approach is better suited for locating true transition states rather than providing a simple estimate of the minimum-energy path. The maximum along the refined ZOOM-NEB-CI pathway corresponds to a genuine first-order saddle point, characterized by a very small imaginary frequency (ca.\ 10$i$ cm$\mathbf{^{-1}}$), whose eigenvector clearly connects the two adsorption motifs. Most importantly, single-point energy refinement of this transition state at the DLPNO-CCSD(T)/aug-cc-pVTZ level, also shown in Figure~\ref{fig:neb_interconversion}, yields an activation energy of only 0.71~\kcalmol\ (357~K) for the H-bond to hemibond conversion. Using this barrier and assuming a pre-exponential factor of $1\times10^{12}$~s$^{-1}$, the lifetime of the H-bonded state is estimated to be 2214~s at 10~K, 870~ps at 50~K, and 7.5~ps at 150~K. These lifetimes clearly indicate that the H-bonded state is rapidly converted on astronomical timescales, implying that its population during the warm-up phase of a molecular cloud collapse is expected to be negligible. At very low temperatures, our calculations also suggest a low population of the H-bonded state. However, it should be noted that the present analysis considers only a single binding site. As shown by \citet{enrique-romero_complex_2024}, hydrogen-bonded configurations exhibit a distribution of binding energies, which may render some sites metastable for timescales long enough to be chemically relevant. Even so, our results unambiguously demonstrate that the hemibonded motif is significantly more abundant than the H-bonded one. This finding underscores the importance of dynamical effects in the binding of interstellar molecules to water ice surfaces and motivates future work explicitly addressing the dynamic behaviour of the CN radical on \ce{H2O} ices.

\section{Eley-Rideal mechanism} \label{app:er}

\begin{figure}
    \centering
    \includegraphics[width=\linewidth]{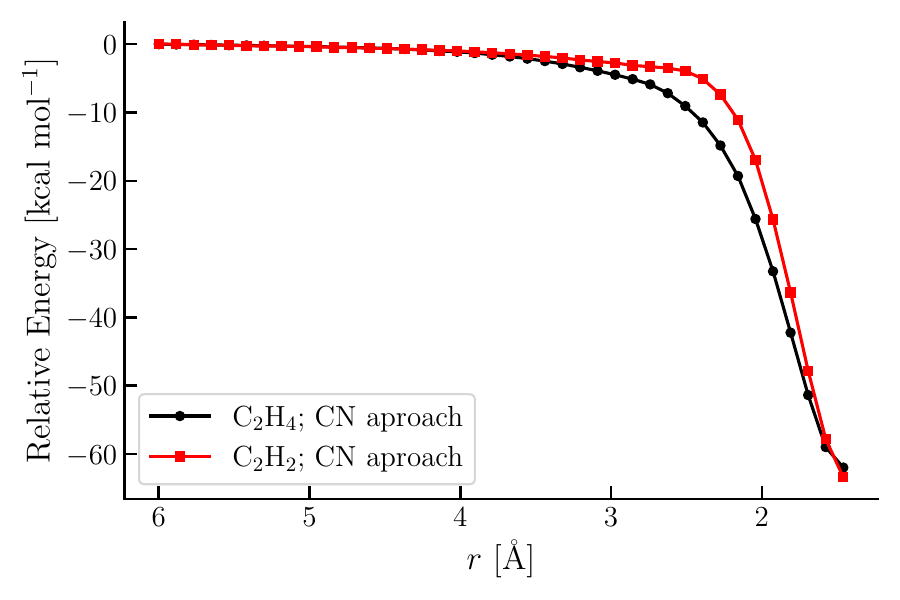}
    \caption{Potential energy scans for the Eley-Rideal approach of an incoming gas-phase CN radical attacking either \ce{C2H2} or \ce{C2H4}. The aftermath of the reaction can be considered analogous to the Langmuir-Hinshelwood mechanism discussed in the main text. The endpoints of the scan represent the radicals \ce{C2H4CN} and \ce{C2H2CN}.}
    \label{fig:er}
\end{figure}

In the main text, we focus on the investigation of the Langmuir--Hinshelwood reactivity of CN radicals with hydrocarbons, which constitutes the dominant mechanism for this reaction in cold to warm icy environments. This is because the surface coverage of the reactants is too low for the alternative Eley--Rideal mechanism to become competitive. Nevertheless, from a conceptual standpoint, it is instructive to explore the Eley--Rideal pathway, as it is not influenced by the water ice matrix and therefore more closely resembles the behaviour expected in the gas phase.

Two possible Eley--Rideal scenarios can be envisaged. In the first, the CN radical approaches a hydrocarbon adsorbed on the surface, while in the second, the hydrocarbon approaches a hemibonded CN species. We discard the latter case, since our scans reveal a clear pre-adsorption of the hydrocarbon on the ice surface. This pre-adsorption provides indirect evidence for the presence of an energy barrier associated with breaking the CN--\ce{H2O} hemibond, reinforcing the electronic nature of this barrier (see the main discussion in Section~\ref{sec:results_ener} and Appendix~\ref{app:c}).

By contrast, the alternative scenario, namely CN attacking a pre-adsorbed hydrocarbon, is found to closely resemble the corresponding gas-phase reaction \cite{balucani_formation_2000, balucani_crossed_2000, huang_crossed_1999,kaiser_formation_2012}, exhibiting a barrierless profile. The results of this scan are shown in Figure~\ref{fig:er}, where the barrierless entrance channel is clearly apparent. The reaction energy inferred from the scan, exceeding $-60~\kcalmol$, is consistent with the fact that, unlike in the diffusive mechanism, an Eley--Rideal scheme includes the contribution from the interaction of the incoming CN radical, specifically its binding energy, in the overall reaction energetics. 

Therefore, astrophysical models including Eley-Rideal chemistry could incorporate the above indicated reactions, with CN coming from the gas phase in their reaction network (and forming \ce{C2H2CN} and \ce{C2H4CN}, see Section \ref{sec:results_ener}), but not the reactions where either \ce{C2H4} or \ce{C2H2} reaction with a pre-adsorbed CN molecule.

\section{Nature of the CN + C$_{\rm X}$H$_{\rm Y}$ reactions on ices} \label{app:c}

\begin{deluxetable}{lccccccc}
\tablecaption{L\"owdin spin population (in a.u. at the M06-2X(D3)/def2-TZVPD level) at the atoms participating in the CN--ice hemibond (carbon atom of the CN radical and oxygen atom of the neighbouring water molecule) and at the two carbon atoms of \cxny, for the reactant and transition states of the reactions shown in Figure~\ref{fig:structures_start}.
\label{tab:partial_charges}}
\tablehead{
\colhead{Reaction} &
\colhead{Orientation} &
\colhead{Spin on C (R)} &
\colhead{Spin on O (R)} &
\colhead{Spin on \cxny\ (R)} &
\colhead{Spin on C (TS)} &
\colhead{Spin on O (TS)} &
\colhead{Spin on \cxny\ (TS)}
}
\startdata
\ce{CN + C2H2 -> C2H2CN} & Parallel   & 0.58 & 0.31 &  0.00    &  0.77    &  0.08    & 0.07     \\
\ce{CN + C2H2 -> C2H2CN} & Collinear  & 0.57  &  0.31  &  0.00    &  0.73    &  0.13   & 0.08    \\
\ce{CN + C2H4 -> C2H4CN} & Parallel   &   0.58   &  0.31    &   0.00   & 0.76  &  0.09    &  0.08    \\
\ce{CN + C2H4 -> C2H4CN} & Collinear  &   0.58   &  0.31   &  0.01  &  0.71    &   0.13   &   0.10   \\
\enddata
\tablecomments{
We note that the spin populations do not have to sum strictly one as other atoms in the \cxny-CN-\ce{14H2O} also have small positive or negative spin populations
}
\end{deluxetable}

%\begin{figure}[htbp]
%\centering
%\includegraphics[width=0.5\linewidth]{react-spin.pdf} \\[2em]
%\includegraphics[width=0.4\linewidth]{ts-spin.pdf}
%\caption{Depiction of spin density at the reactant (left) and transition state (right), the reaction with \ce{C2H4} in the parallel orientation is selected as an example. The isovalue is 0.005~a.u.}
%\label{fig:spin}
%\end{figure}

A crucial aspect in the modeling of the reactions investigated in this work is the nature of the kinetic barrier that emerges on ices. If the barrier is purely diffusive its influence must be incorporated in the $\left( \dfrac{1}{\tau_{\rm C_{X}H_{Y}}} + \dfrac{1}{\tau_{\rm CN}} \right)$ term of Equation \ref{eq:lh} with $\epsilon_r$=1.0. Otherwise, $\epsilon_r$ must be explicitly calculated according to equation \ref{eq:efficiency}, with k$_{r}$ exponentially depending on the height of the barrier in a Boltzmann picture (Equation \ref{eq:k}) and competing with diffusion out of the reaction center. In the latter case, we refer to the barriers as ``electronic'' as they arise from purely electronic effects between the adsorbates and the ice matrix.

To the best of our knowledge, there is not abundant exploration to the difference between diffusive and electronic barriers in radical-radical reactions in the literature, with a small discussion already presented in \citet{postulka_diffusive_2025}. In general, it is difficult to establish the true nature of the reactions, but in this case, the analysis is simpler, owing to the hemibond between CN and \ce{H2O}, that allows to track the evolution of the spin density during the reaction. The spin population very clearly determines the nature of the reaction, as there is an evident spin density migration from the O-atom of the hemibonded water to the CN radical at the TS. This is clear evidence of the neccesity of breaking the hemibond and evinces the electronic nature of the barrier. In this work, we determine the nature of the reaction by inspecting the spin population at the reactant and transition states, specifically using a L\"owdin pouplation analysis whose result are gathered in Table \ref{tab:partial_charges}. An inspection of the Table shows that, for all inspected cases, the reactant state has a lower spin population on the hemibonded C with respect to the transition state. Such electron migrations are not expected in purely diffusive mechanisms hence demonstrating the electronic nature of the barrier and the need of using equation \ref{eq:efficiency} to estimate the reaction efficiency.

%\section{Raw values of the CN + \cxny\ reaction rate constants} \label{app:d}

%\begin{figure}
%    \centering
%    \includegraphics[width=\linewidth]{rates_2x2_square.pdf}
%    \caption{Reaction rate constants for the addition of CN to \cxny\ and used in Section \ref{sec:kinetic}. The values of the rate constants used in this plot can be retrieved as data behind the figure.}
%    \label{fig:raw_rates}
%\end{figure}

%We report the values of the reaction rate contants $k_{r}$ used to generate the plots in Figure \ref{fig:competition_c2h4} and \ref{fig:competition_c2h2} as data behind the tables in Figure \ref{fig:raw_rates}.

%Analysis of the TS wavefunction

%% For this sample we use BibTeX plus aasjournalv7.bst to generate the

\bibliography{sample701}{}
\bibliographystyle{aasjournalv7}



\end{document}